\documentclass[%
 aip,
 amsmath,amssymb,
 reprint,%
]{revtex4-1}

\usepackage{graphicx}
\usepackage{dcolumn}
\usepackage{bm}

\usepackage[utf8]{inputenc}
\usepackage[T1]{fontenc}
\usepackage{mathptmx}

\usepackage{color}

\begin{document}

\preprint{AIP/123-QED}

\title{Tip-induced domain protrusion \\in ferroelectric films with in-plane polarization}

\author{S. Kondovych}
 \email{svitlana.kondovych@gmail.com}
 \affiliation{Laboratory of Condensed Matter Physics, University of Picardie, 33 rue St. Leu, Amiens 80039, France}
 \affiliation{Life Chemicals Inc., Murmanska st. 5, Kyiv, 02660, Ukraine}
\author{A. Gruverman}%
\affiliation{ Department of Physics and Astronomy, University of Nebraska, Lincoln, NE 68588-0299, USA}%
\author{I. Luk'yanchuk}
 \affiliation{Laboratory of Condensed Matter Physics, University of Picardie, 33 rue St. Leu, Amiens 80039, France}
\affiliation{Faculty of Physics, Southern Federal University, 5 Zorge Str., 344090, Rostov-on-Don, Russia}


\begin{abstract}
Charge manipulation and fabrication of stable domain patterns in ferroelectric materials by scanning probe microscopy open up broad avenues for the development of tunable electronics. Harnessing the polarization energy and electrostatic forces with specific geometry of the system enables producing the nanoscale domains by-design. Along with that, domain engineering requires mastery of underlying physical mechanisms that govern domain formation. Here, we present a theoretical description of the domain formation by a scanning probe microscopy tip in a ferroelectric film with strong in-plane anisotropy of polarization. We demonstrate that local charge injection produces wedge-shaped domains that propagate along the anisotropy axis, whereas the tip-written lines of charge generate a comb-like domain structure.   
The results of our calculations agree with earlier experimental observations and allow for the optimization of the targeted domain structures.
\end{abstract}

\maketitle


\section{introduction}

Anisotropy of ferroelectrics and their high sensitivity to external electric fields allow for effective control of their functional properties for successful implementation of novel ferroelectric technologies\cite{Dawber2005,Scott2007}, such as data storage \cite{scott2013book}, non-binary and neuromorphic computing systems\cite{baudry2017ferroelectric,li2020reproducible,zhong2020flexible}, terahertz emitters and detectors\cite{luk2018electrodynamics}, low-dissipate computing circuits with negative capacitance\cite{khan2015negative,iniguez2019ferroelectric,luk2019harnessing}. A great variety of ferroelectric polarization topological structures\cite{seidel2016topological}, including domains and domain walls (DW) \cite{tagantsev2010domains}, vortices\cite{naumov2004unusual,lahoche2008stability,yadav2016observation}, skyrmions\cite{nahas2015discovery,das2019observation,tikhonov2020controllable}, and hopfions\cite{luk2020hopfions} makes
textured ferroelectric materials attractive not only for fundamental studies but also for industrial applications in nanoelectronics\cite{seidel2019nanoelectronics}. 
 Properties of the ferroelectric films critically depend on the emerging domain patterns, which ultimately determine the resulting operation modes of ferroelectric-based electrically-controllable devices. The decisive role here belongs to the unique tunability of the ferroelectric film functionalities by the external parameters such as strain-induced anisotropy, temperature, sample geometry, mechanical stress, and applied voltage. This makes the fabrication of stable domain textures with tailored geometry an essential goal of device design and engineering\cite{catalan2012domain}.

Whereas the out-of-plane domain structures and corresponding switching phenomena in thin ferroelectric crystals, films, and superlattices have been extensively investigated both theoretically\cite{Bratkovsky2000,kornev2004ultrathin,mokry2004size,stephanovich2005domain,aguado2008,luk2009universal} and experimentally\cite{streiffer2002observation,zubko2010x,hruszkewycz2013imaging,lukyanchuk2014high,zubko2016negative}, the statics and, particularly, dynamics of in-plane ferroelectricity have been studied to a much lesser extent. 
This is unfortunate as the investigation of in-plane polarization reversal provides an invaluable insight into the process of the forward (along the polar axis) domain growth, which is difficult to attain in conventional ferroelectric switching studies\cite{kagawa2014polarization}. Recent experimental studies performed by visualization of the domain structure on the non-polar surface allowed for comparative analysis of the forward and lateral growth kinetics, direct assessment of the role of structural defects in domain motion, and effect of screening on the kinetics and stability of the charged domains walls\cite{sluka2012enhanced,ievlev2015symmetry,Lu2015,turygin2018self,bak2020observation,sharma2013nanoscale,whyte2014ferroelectric,whyte2015diode}. Most of these studies have been performed by means of piezoresponse force microscopy (PFM), where domain manipulation on the non-polar surface was realized due to the lateral components of the electric field generated by a PFM probe. Further expansion of this approach to the in-plane polarization control requires a better understanding of the fundamental mechanisms driving the domain growth in the conditions of a highly non-uniform electric field and strongly anisotropic medium.

\begin{figure}[h!]
\includegraphics[width=0.45\textwidth]{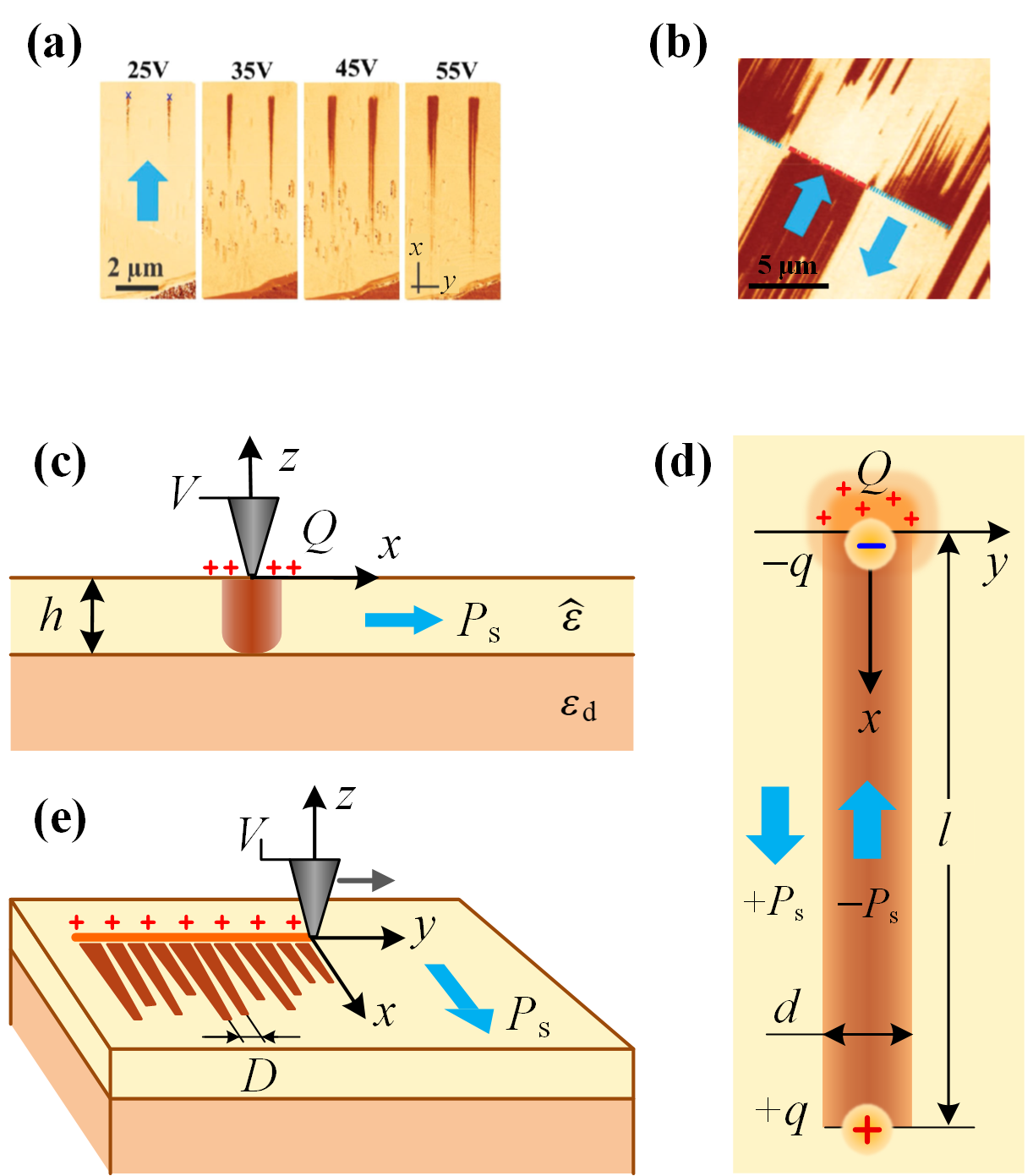}%
\caption{\label{fig:model} Wedge-shape domain formation in the ferroelectric film with uniaxial in-plane anisotropy. (a) and (b) Experimental domains\cite{Lu2015} generated by PFM tip in a planar organic DIPA-B ferroelectric slab under the different applied voltages for point-localized and linear distribution of the tip-injected charges respectively (reprinted with permission from Adv. Mater. 27, 7832 (2015). Copyright 2015 John Wiley and Sons). 
The yellow and brown colors correspond to the domains with the opposite orientation of polarization vectors (blue arrows).
(c) Cross-section view of the domain in the film. The tip-generated voltage pulse $V$ injects localized positive charge $Q$ into the ferroelectric film of thickness $h$ with in-plane spontaneous polarization $P_{\rm s}$ and dielectric tensor $\hat{\varepsilon}$. The film is deposited on the dielectric substrate with dielectric constant $\varepsilon_{\rm d}$. The $x$-axis of the coordinate system is directed along $P_{\rm s}$, and the $z$-axis is perpendicular to the film surface. (d) View from the top. The injected charge $Q$ attracts the negative depolarization charge $-q$ and repulses the positive one $+q$, resulting in the appearance of an elongated domain of reversed polarization, $-P_{\rm s}$, of the length $l$ and width $d$. (e) The linear charge injected by the moving along the $y$-axis tip extrudes a multiple-domain comb-like structure with the period $D$.}
\end{figure}

In this paper, we present a theoretical consideration on the earlier experimental results on the PFM-induced formation of the in-plane wedge-shaped domains in uniaxial ferroelectric crystal diisopropylammonium bromide (DIPA-B) \cite{Lu2015}. 
Modeling the tip-induced voltage in the experimental setup as a point-like and linear charge injection, we explain the origin of a single wedge-shaped and comb-like domain structures that were observed (See  Fig.\,\ref{fig:model}a and Fig.\,\ref{fig:model}b, respectively) and calculate their parameters. Qualitative and quantitative agreement of the obtained results with the experiment enables optimization and by-design engineering of the domain patterns in ferroelectric films with in-plane anisotropy.

\section{Model}

\subsection{Geometry and formation of the domain structure}

The principal parameters of the system are illustrated in Fig.\,\ref{fig:model}c. We consider a ferroelectric film of thickness $h$ with the in-plane spontaneous polarization directed along the x-axis, $\mathbf{P}=(P_{\rm s},0,0)$, deposited on the dielectric substrate with the dielectric constant $\varepsilon_{\rm d}$.
The uniaxial dielectric tensor of the film has the diagonal components $\hat\varepsilon=(\varepsilon_{11},\varepsilon_{22},\varepsilon_{33})=(1,\eta^2,\eta^2)\varepsilon_{11}$, where $\eta^2>1$ is the anisotropy factor.

A tip-applied voltage pulse injects charge $Q$, positive for definiteness, into the sample and generates an elongated wedge-like domain as shown in Fig.\,\ref{fig:model}d (view from the top). The domain grows along the $x$-axis, acquiring the width $d$ and length $l$. The depolarization charges due to polarization reversal at the tip-application point and at the domain ending are equal respectively to $\mp q\simeq \mp 2(hd) P_{\rm s}$.
The dynamics of the domain evolution after charge $Q$ injection is governed by the ponderomotive force acting on the terminal charge $+q$ from the charge $Q-q$ at the domain origin, composed from the injected charge $Q$ and the depolarization charge $-q$. 
In case $Q>q$ the interaction has the repulsive character that pushes the domain termination point away from the tip-induced charge $Q$, resulting in protrusion of the domain in the plane of the film until the increasing energy of the DW compensates the electrostatic force. This defines the equilibrium domain length.  
We take the domain width proportional to the effective charge at the domain origin\cite{lilienblum2011determination,li2012piezoresponse,Lu2015} and present this dependence as  
$\varkappa_1 hP_{\rm s }d=Q-q$, where $\varkappa_1$ is the material-dependent proportionality coefficient to be found from the experimental data.

Another experimental result is related to the in-plane domain formation by moving the electrically-biased PFM tip over the non-polar surface of the DIPA-B sample in the direction perpendicular to the polar axis injecting a linear charge with density $\lambda_Q$. This stimulates the partial polarization reorientation at one semi-space with the formation of the compensating depolarization charge due to the head-to-head (or tail-to-tail) polarization junction at the line. Such reorientation causes the protrusion of a number of charges into the material and generates a comb-like domain pattern with the period $D$ as shown in Fig.\,\ref{fig:model}e.  
The effective charge density at the line can be estimated as $\lambda_Q-\lambda_{q}$, where $\lambda_{q}=2 \alpha P_{\rm s} h$ is the linear depolarization charge due to polarization re-orientation in domains, and parameter $\alpha\simeq d/D<1$ reflects the relative weight of the domain regions along the tip-written line. The initial domain formation occurs in a critical way to overcome the initial potential barrier of nucleation. Hence, the width of the created domain is of the order of the nucleus critical size that, as indicated by experiments\cite{Lu2015}, is proportional to $\lambda_Q-\lambda_{q}$. We assume that $\varkappa_2 hP_{\rm s }d=(\lambda_Q-\lambda_{q})D$ with experimentally-defined coefficient $\varkappa_2$.

Although the mechanism of domain formation is similar for both cases, the quantitative parameters of domains depend on the dimensionality of the injected charge regions, as will be shown further.

\subsection{Total energy of the system}

The optimal geometric parameters of the domain structure are obtained by minimization of the total energy of the system that includes the electrostatic energy and the DW energy: 
\begin{equation}
\mathcal{W}_{\rm tot} = \mathcal{W}_{\rm el}+\mathcal{W}_{\rm dw}. \label{total_energy}
\end{equation}
The DW energy, $\mathcal{W}_{\rm dw}$, of the single domain shown in Fig.\,\ref{fig:model}d is written as $2\sigma_{\rm dw}hl$, where the surface tension of the DW of thickness $\xi\simeq 1$nm is estimated as\cite{lukyanchuk2014high} $\sigma_{\rm dw} \simeq \xi  P_{\rm s}^2 / \varepsilon_0\varepsilon_{11}$. Here, $\varepsilon _{0}$ is the vacuum permittivity, and factor 2 accounts for two DWs forming the domain. In the case of the comb-like domain structure, the DW energy per unit of length along the charged line is expressed as $2\sigma_{\rm dw}{hl}/{D}$.

The electrostatic energy $\mathcal{W}_{\rm el}$ describes the interaction of the charges $+q$, located at the terminal point of the domains with the tip-induced charges at the domain origin, $Q-q$ for the case of the localized charge and $\lambda_Q-\lambda_{q}$ for the charged line. 
We further derive this energy for the cases of the point-localized and linear charge injection to find the equilibrium domain parameters.

\subsection{Electrostatic energy of the single domain}

\begin{figure}[h!]
\centering
\includegraphics[width=0.49\textwidth]{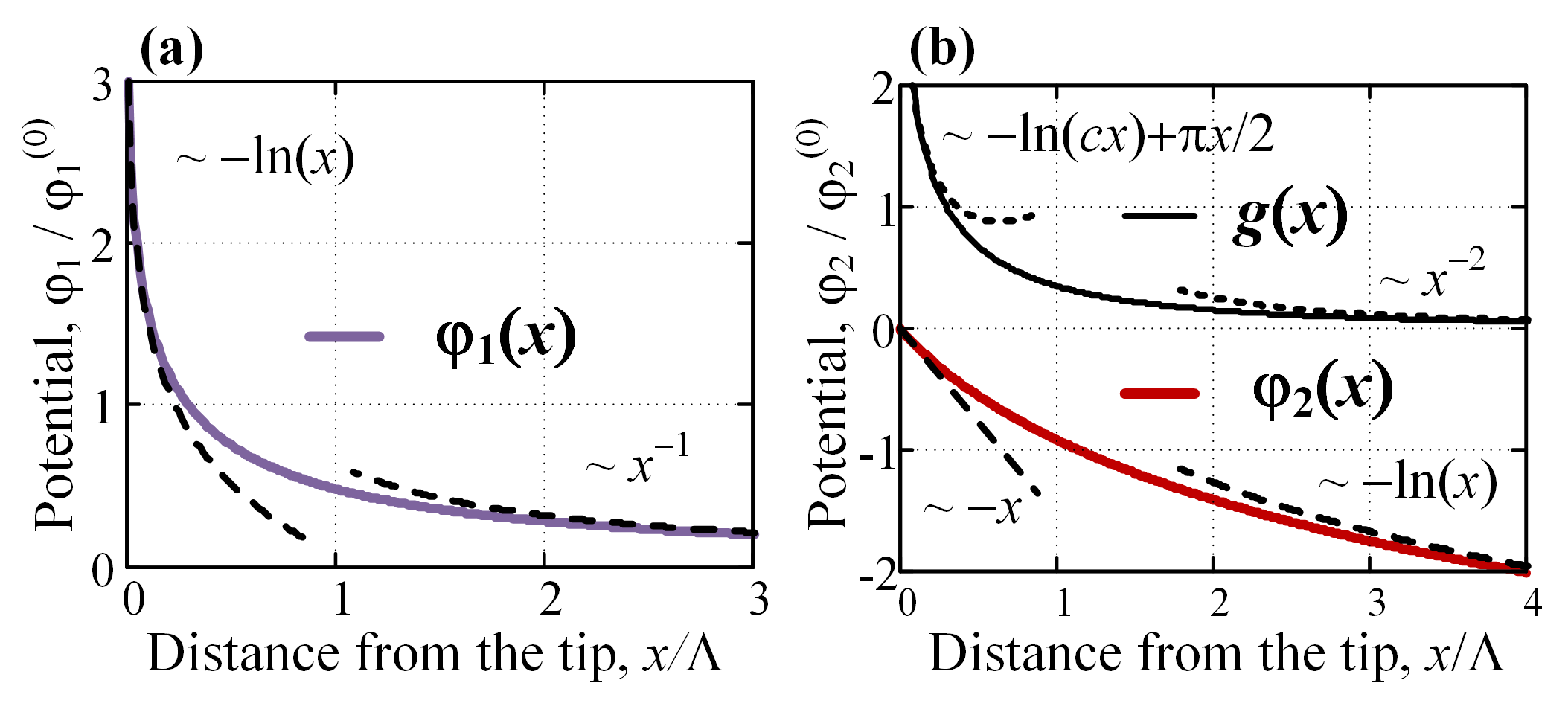}
\caption{\label{fig:potential} Electrostatic potential of the tip-generated field in the ferroelectric film. (a) The purple line corresponds to the space variation of the potential, $\varphi_1(x)$, created by the point-localized tip-injected charge. (b) The potential, $\varphi_2(x)$, induced by the tip-drawn linear charge (red line). The black line represents the auxiliary function $g(x)$, which forms the expression (\ref{pot_2d}) for $\varphi_2(x)$.   
The distance $x$ in both plots is scaled in units of the characteristic length $\Lambda$ and potentials are scaled in units of characteristic potentials, $\varphi_1^{(0)}$ in (a) and $\varphi_2^{(0)}$ in (b). The dashed lines depict the limit cases of the functional behavior at $x\gg\Lambda$ and at $x\ll\Lambda$, as described in the text.}
\end{figure}

According to electrostatics, the energy of the interaction between charges $+q$ and $Q-q$ is written as $\mathcal{W}_{\rm el}=q\varphi_1(x)$, where $\varphi_1(x)$ is the potential of the charge $Q-q$ at the location $x$ of the charge $q$. This potential is found as a solution of the Poisson's equation with appropriate boundary conditions (see Appendix A for details). At distances larger than the film thickness, $x\gg h$, the dependence $\varphi_1(x)$ is written as:
\begin{equation}
\varphi_1(x) \simeq \varphi_1^{(0)}\Phi _{0}\left(\frac{x}{\Lambda }\right). \label{pot_3d}
\end{equation}
Here, $\Phi _{0}(\zeta)=H_{0}\left( \zeta\right) -N_{0}\left( \zeta\right)$ denotes the difference between the zero-order Struve and Neumann functions \cite{Abramowitz}, and the $\Lambda$ and $\varphi_1^{(0)}$ are the characteristic length and characteristic potential, respectively:  
\begin{equation}
\Lambda \simeq  \frac{\sqrt{\varepsilon_{11}\varepsilon_{33}}+\varepsilon _{\rm d} }{
\varepsilon _{\rm d} }\eta h,  \qquad \varphi_1^{(0)}=\frac{Q-q}{4\varepsilon_0\varepsilon_{33}\eta h}, \label{lambda}
\end{equation}
where the expression for $\Lambda$  is given for $\varepsilon_{11},\varepsilon_{33},\varepsilon_{\rm d} \gg 1$, for arbitrary dielectric constants see Eq.(\ref{lambda_general}) in Appendix A.
This characteristic length separates the regions with different types of the electrostatic interaction\cite{kondovych2017gate}. Namely, for $x\ll\Lambda$, the interaction manifests the 2D logarithmic behavior, $\varphi_1(x) \sim -\ln (x/\Lambda)$, whereas for $x\gg\Lambda$ the 3D Coulomb decay of the potential is obtained, $\varphi_1(x) \sim \Lambda/x$, see Fig.\,\ref{fig:potential}a. The characteristic potential $\varphi_1^{(0)}$ estimates the value of the potential at $x \sim \Lambda$.

\subsection{Electrostatic energy of the comb-like domain structure }

The electrostatic energy (per unit of length along the charged line) of the interaction of the depolarization charges located at the termination of the comb-like domains with the tip-induced linear charge is written as  $\mathcal{W}_{\rm el}=\lambda_q\varphi_{2}$, where the electrostatic potential $\varphi_{2}$ is due to the field created by the DW. This potential at $x \gg h$ is expressed as: 
\begin{equation}
 \varphi_{2}(x)  \simeq \varphi_2^{(0)} \left[ \ln \left(c\frac{x}{\Lambda }\right)+g\left( \frac{x}{%
\Lambda }\right) \right], \quad \varphi_2^{(0)}=\frac{\lambda_q-\lambda_Q}{\pi \varepsilon _{0}\varepsilon _{\rm d}}, \label{pot_2d}
\end{equation}
where $\ln c \simeq 0.577$ is the Euler's constant, 
the function  $g\left( \zeta\right) =\left( {\pi }/{2}-\mathrm{%
Si\,}\zeta\right) \sin \zeta-\mathrm{Ci\,}\zeta\cos \zeta$ is  the auxiliary 
trigonometric integral function\cite{Abramowitz}, the characteristic length $\Lambda$ is given by  Eq.\,(\ref{lambda}), and $\varphi_2^{(0)}$ is the characteristic potential of the system at $x\sim \Lambda$.
At small distances, $x \ll \Lambda$, the $g$-function behaves as $g(\zeta)\simeq -\ln(c\zeta)+\pi \zeta/2$, and the resulting potential has the linear behaviour. 
At large distances, $x \gg \Lambda$, the vanishing behaviour of $g(\zeta) \simeq -1/\zeta^2$ results in the logarithmic decay of the potential (Fig.\,\ref{fig:potential}b).

\section{Results}
\subsection{Single domain}

The equilibrium length, $l$, of the single domain is calculated by the minimization of the total energy $\mathcal W_{\rm tot}$ (Eq.\,(\ref{total_energy})) with respect to $x$. The final expression, relating $l$ and the domain width $d$, which depends on the injected charge, is written as 
\begin{equation}
\label{l_opt_3d}
\Phi_{-1}\left(\frac{l}{\Lambda }\right)=-\frac{4\Lambda\xi}{\varkappa_1 \eta d^2},
\end{equation}
where $\Phi_{-1}(\zeta)\equiv d\Phi_{0}/d\zeta$. The dependence $l(d)$ in limit cases, which correspond to different regimes of domain growth, is found explicitly from Eq.(\ref{l_opt_3d}) using the asymptotic expressions for  $\Phi_{0}(\zeta)$. The equilibrium domain length is estimated as $l \simeq (\varkappa_1 \eta/2\pi\xi)d^2 $ for intermediate lengths $h\ll l \ll \Lambda $. 
For the longer domains with $l \gg \Lambda$, it is given by 
$l \simeq ({\varkappa_1 \eta \Lambda }/{2 \pi \xi})^{1/2}d $.

\subsection{Comb-like domain structure}

Minimization of $\mathcal W_{\rm tot}$ in the case of multiple-domain comb-like structure gives implicitly the dependence $l(d)$, 
\begin{equation}
f\left( \frac{l}{\Lambda}\right) 
=\frac{\pi \varepsilon_{\rm d}\xi \Lambda D}{\varkappa_2 \varepsilon_{11} h d^2},
\label{l_opt}
\end{equation}
where
$f\left( \zeta\right) $\ is
related to $g(\zeta)$ as\cite{Abramowitz} $f\left( \zeta\right)=g'\left( \zeta\right)+1/\zeta$. 
This function decreases monotonously and for the large values of argument is simplified as $f(\zeta) \simeq 1/\zeta$. We obtain that for large $l \gg \Lambda $, the equilibrium domain length depends 
on the domain width $d$ as: $l \simeq (\varkappa_2 \varepsilon_{11} h/\pi \varepsilon_{\rm d}\xi D)d^2 $.

\section{Discussion}

\begin{figure}[h!]
\centering
\includegraphics[width=0.48\textwidth]{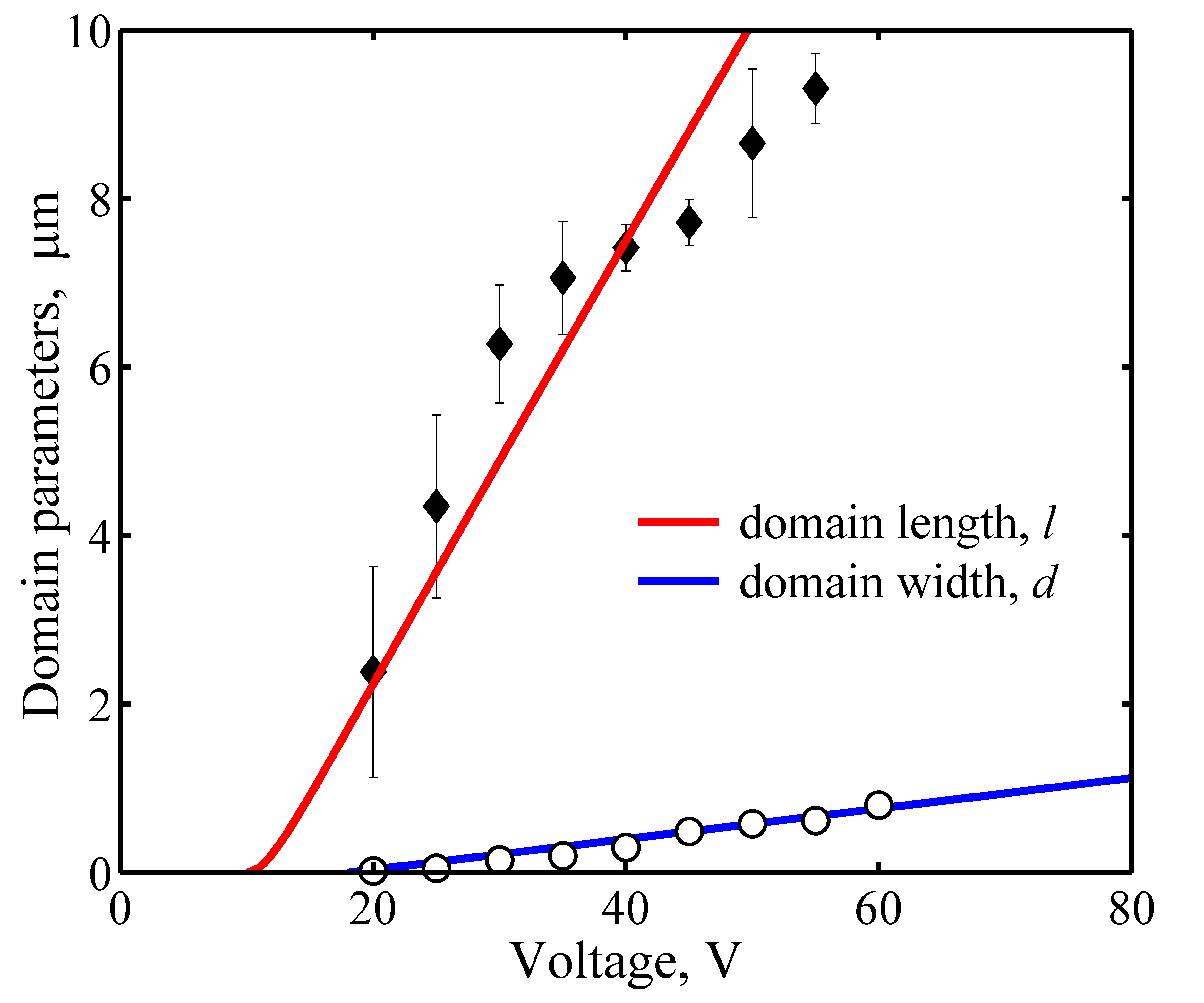}
\caption{\label{fig:fit} Length and width of the tip-induced single domains as a function of the applied voltage. The experimental data for the domain length and width in the DIPA-B slab\cite{Lu2015} are denoted by the diamond and circle markers respectively.
The blue line corresponds to the best linear fit of the experimental domain width $d$.
The red line refers to the calculated domain length $l$.}
\end{figure}

The obtained results allow not only for the qualitative description of the formation mechanism of the wedge-shaped domains in the uniaxial ferroelectric films but also for the quantitative comparison with the available experimental data.  Fig.\,\ref{fig:fit} demonstrates the dependence of the geometrical parameters of the string domains created by the local application of the voltage pulse with amplitude  $V$, which injects a localized charge $Q$, proportional to $V$\, \cite{silva2010charge}. 
In this figure, we compare the experimental data for length and width of the shown in Fig.\,\ref{fig:model}a single domain in DIPA-B slab of thickness $h=150$nm with our theoretical results. The red line presents the calculated dependence $l(d)$ described by  Eq.\,(\ref{l_opt_3d}) with $P_s = 23$ $\mu$C/cm$^2$, $\varepsilon_{11}=80$, $\varepsilon_{33}=40$, $\varkappa_1\simeq 1$, and $\xi \simeq 1$nm. The characteristic length $\Lambda$ is estimated from (\ref{lambda}) with $\varepsilon_{\rm d}\simeq 12$ (for Si substrate) and  $\eta=(\varepsilon_{11}/\varepsilon_{33})^{1/2}\simeq 1.4$ as $\Lambda \simeq 1\mu$m. 
The domain width $d$ at given voltage was taken from the best linear fit of the experimental data (the blue line). 

Note that the more precise variational approach would account for the spatial distribution of the depolarization charge over the entire domain boundary, revealing the wedge form of domains, similar to that observed in the experiment (Fig.\,\ref{fig:model}a).  On the quantitative level, this results in the dimensionless correction coefficient that does not affect the obtained functional dependence of the domain parameters on the voltage, given in Fig.\,\ref{fig:fit}.

 Our calculations demonstrate a very good match with the experimental data and can be used as the base for the design and optimization of the tip-generated domains in ferroelectric films with in-plane anisotropy. In addition, since the domain length appears to be larger than the characteristic length $\Lambda$, the derived above approximate expression $l \simeq ({\varkappa_1 \eta \Lambda }/{2 \pi \xi})^{1/2}d $ can be employed for practical use.

Estimation of parameters of the domain comb-like structure generated by the moving PFM tip has a more qualitative character. The challenge here is to properly take into account the interaction between the charges at the terminal points of domains. This collective effect may not only affect the quantitative estimations given by Eq.\,(\ref{l_opt}) but also lead to the instability of the alignment of the terminal charges in a line. The repulsion between these charges will result in the variable length of the domains, as seen in Fig.\,\ref{fig:model}b. However, the order-of-magnitude estimation of the characteristic (average) domain length gives $l \simeq (\varkappa_2 \varepsilon_{11} h/\pi \varepsilon_{\rm d}\xi D)d^2 \simeq 8\mu$m with $d/D \simeq 0.5$, $\varkappa_2 \simeq 1$, and $d \sim 50$nm for the charged line drawn at $85 V$, which provides a satisfactory agreement with the experiment (Fig.\,\ref{fig:model}b).

\begin{acknowledgments}
This work was supported by the H2020-MSCA-RISE actions ENGIMA (Grant No: 778072) and MELON (Grant No: 872631).
We also acknowledge the assistance of I.F. Lab and Mileage Mobility Action.
\end{acknowledgments}

\section*{DATA AVAILABILITY}
Data sharing is not applicable to this paper as no new data were created or analyzed in this study.

\appendix
\section{Methods}
The electrostatic potential created by a point-like charge $Q$ inside the ferroelectric  layer (Fig.\,\ref{fig:model}c) is found as a solution of Poisson's equation, 
\begin{equation}
   \eta \partial _{x}^{2}\varphi+\partial _{y}^{2}\varphi+\partial _{z}^{2}\varphi=-\frac{Q}{\varepsilon _{0}\varepsilon_{33}} \delta_3(x,y,z) ,
\end{equation}
 with boundary conditions at the ferroelectric interfaces:
 \begin{eqnarray}
z &=& 0 : \quad \varphi =\varphi_{+}, \quad \varepsilon_{33}\partial _{z}\varphi =\partial _{z}\varphi_{+}; 
\label{BC1} 
\\
z&=&-h : \quad \varphi =\varphi_{-}, \quad \varepsilon_{33}\partial _{z}\varphi =\varepsilon_{\rm d}\partial _{z}\varphi_{-}. \, \label{BC2}   
\end{eqnarray}%
 Here, $\delta_3(x)$ is the Dirac delta-function, $\varphi_{\pm}$ denote the electrostatic potentials in the regions above and below the film, respectively; $\nabla^{2}\varphi_{\pm}=0$. We search for the solution in form:
\begin{equation}
\varphi(x,y,z) \sim \int_0^\infty \int_0^\infty e^{\pm z \sqrt{k_y^2+\eta^2k_x^2}}\cos{(k_xx)}\cos{(k_y y)} dk_x dk_y.
\end{equation}
At the film surface, $z=0$, at distances larger than the film thickness, $x\gg h$, we obtain the expression for the electrostatic potential ${\varphi (x,y)}$: 
\begin{equation}
\varphi(x,y) \simeq \frac{Q-q}{4\varepsilon_0\varepsilon_{33}\eta h}\Phi _{0}\left(\frac{\sqrt{x^2+\eta^2 y^2 }}{\Lambda }\right),
\end{equation}
that gives Eq.\,(\ref{pot_3d}) at $y=0$. In general case of arbitrary values of dielectric constants, the characteristic length, $\Lambda$, is given by: 
\begin{equation}
\Lambda =\frac{\left( \varepsilon _{a}+\sqrt{\varepsilon_{11}\varepsilon_{33}}\right)\left( \varepsilon _{b}+\sqrt{\varepsilon_{11}\varepsilon_{33}}\right) }{\varepsilon_{33}\left(
\varepsilon _{a}+\varepsilon _{b}\right) }h, \label{lambda_general}
\end{equation}
which in case of high permittivity materials with $\varepsilon_{11},\varepsilon_{33},\varepsilon_{\rm d} \gg 1$ reduces to (\ref{lambda}).

In a similar way, for the charged line confined in the film at $-h<z<0$ with the linear charge density $\lambda_Q$ along $y$ axis, the solution of Poisson's equation,
\begin{equation}
   \eta \partial _{x}^{2}\varphi+\partial _{z}^{2}\varphi=-\frac{\lambda_Q h}{\varepsilon _{0}\varepsilon_{33}} \delta (x) ,
\end{equation}
with the boundary conditions (\ref{BC1})-(\ref{BC2}) results in Eq.\,(\ref{pot_2d}), see Ref.\cite{kondovych2017nondestructive} for details. 


\providecommand{\noopsort}[1]{}\providecommand{\singleletter}[1]{#1}%

\end{document}